\documentclass[prb,superscriptaddress,preprint,showpacs,unsortedaddres
s]{revtex4}

\usepackage{graphicx}
\usepackage{dcolumn}
\usepackage{bm}

\begin{document}

\title{Samarium Magnetism Studied on SmPd$_{2}$Al$_{3}$ Single Crystal}

\author{J. Posp\'{\i}\v{s}il}

\email{jiri.pospisil@centrum.cz}

\affiliation{Charles University, Faculty of Mathematics and
Physics, Department of Condensed Matter Physics, Ke~Karlovu~5, 121
16 Prague 2, The Czech Republic}

\author{M. Kratochv\'{\i}lov\'{a}}

\affiliation{Charles University, Faculty of Mathematics and
Physics, Department of Condensed Matter Physics, Ke~Karlovu~5, 121
16 Prague 2, The Czech Republic}

\author{J. Prokle\v{s}ka}

\affiliation{Charles University, Faculty of Mathematics and
Physics, Department of Condensed Matter Physics, Ke~Karlovu~5, 121
16 Prague 2, The Czech Republic}

\author{M. Divi\v{s}}

\affiliation{Charles University, Faculty of Mathematics and
Physics, Department of Condensed Matter Physics, Ke~Karlovu~5, 121
16 Prague 2, The Czech Republic}

\author{V. Sechovsk\'{y}}

\affiliation{Charles University, Faculty of Mathematics and
Physics, Department of Condensed Matter Physics, Ke~Karlovu~5, 121
16 Prague 2, The Czech Republic}

\date{\today}

\begin{abstract}
In this paper, specific features of Sm magnetism in an
intermetallic compound have been studied. For this purpose, a high
quality single crystal of SmPd$_{2}$Al$_{3}$ was grown and
subjected to detailed measurements of specific heat,
magnetization, AC susceptibility and electrical resistivity with
respect to temperature and a magnetic field applied along the
principal crystallographic directions. SmPd$_{2}$Al$_{3}$
magnetism was found to be strongly anisotropic with the
easy-magnetization direction along the \textit{c}-axis where the
main magnetic features are concentrated. The \textit{a}-axis
response remains weak, paramagnetic-like, even in the magnetically
ordered state. Ferromagnetism with $\textit{T}$$_{C}$ = 12.4 K has
been indicated by all the measured physical properties. At lower
temperatures, three successive order-order phase transitions have
been observed on the temperature dependence of the specific heat
as three anomalies: at 3.4, 3.9, 4.4 K, respectively. The low
temperature magnetization data can be understood within a scenario
that considers the antiferromagnetic ground state as being
gradually destroyed through a series of four metamagnetic
transitions at 0.03, 0.35, 0.5 and 0.75 T, as detected in the
1.8-K magnetization data. The temperature dependence of the
paramagnetic susceptibility below 200 K can be in the first
approximation interpreted in terms of a Curie-Weiss law modified
by temperature independent van Vleck contribution due to the
low-lying first excited multiplet \textit{J} = 7/2 being
populated. At higher temperatures, involvement of the second
excited multiplet \textit{J} = 9/2 should also be considered. The
experimental data are discussed together with the results of
electronic-structure and crystal-field calculations from first
principles, which were performed as an important part of the study
for comprehension and explanation of the observed behavior of the
SmPd$_{2}$Al$_{3}$ compound.
\end{abstract}

\maketitle

\section{Introduction}
The localized $\textit{4f}$-electron magnetism of lanthanide ions
is by rule characterized by stable magnetic moments, which reflect
the population of energy levels of the ground-state multiplet
determined by total angular momentum $\textit{J}$ and split by
crystal field (CF) interaction. Paramagnetic susceptibility
$(\chi)$ well above the magnetic-ordering temperature obeys the
Curie Weiss law. This behavior may be modified at lower
temperatures in the case of a considerable CF interaction. Sm
represents an exception to this rule: Sm magnetism is not governed
only by the ground state multiplet $\textit{J}$ = 5/2, since the
first (and second) excited multiplet $\textit{J}$ = 7/2
($\textit{J}$ = 9/2) has only an energy of 0.1293 eV (0.2779 eV)
\cite{01}. Therefore the temperature dependence of paramagnetic
susceptibility at temperatures up to about 400 K is influenced by
populating the CF split levels of the first and second excited
multiplets via the temperature-independent Van Vleck terms
\cite{02} and does not obey the Curie-Weiss law. A typical
1/$\chi$ vs. \textit{T} plot for a Sm compound is strongly
nonlinear and usually exhibits a maximum around 400 K, which is
very sensitive to CF and exchange interaction \cite{03}. Since the
samarium magnetic moment at low temperatures is small
($\textit{gJ}$ = 0.71 $\mu$$_{B}$ for the Sm$^{3+}$ free ion)
exchange interactions can be very peculiar and several magnetic
phases may appear in the same material at different temperature
and magnetic-field intervals, respectively. In the present paper
we demonstrate some specific features of Sm magnetism from the
results of our recent study of a high-quality single crystal of
the compound SmPd$_{2}$Al$_{3}$. We focused our attention on
measurements of magnetization ($\textit{M}$), AC susceptibility
($\chi$$_{AC}$), specific heat ($\textit{C}$$_{p}$) and electrical
resistivity ($\textit{$\rho$}$) as functions of temperature
($\textit{T}$) and the external magnetic field ($\textit{B}$). The
anisotropy of magnetic properties has been determined from
measurements in magnetic fields applied along the principal
crystallographic directions [100] and [001]. Along with our
experimental work, we performed various theoretical calculations
of the electronic structure and CF parameters and compared
calculated results with experimental findings. SmPd$_{2}$Al$_{3}$,
as well as its $\textit{RE}$Pd$_{2}$Al$_{3}$ ($\textit{RE}$ = La,
Ce, Nd, Gd) counterparts, adopt the same hexagonal
PrNi$_{2}$Al$_{3}$-type crystal structure \cite{04}. It is worth
noting that the U isostructural analog, UPd$_{2}$Al$_{3}$, is a
well-known heavy-fermion antiferromagnetic superconductor
($\gamma$ = 150 mJ K$^{-2}$mole$^{-1}$, $\textit{T}$$_{C}$ = 2 K,
$\textit{T}$$_{N}$ = 14 K) \cite{05}. The nonmagnetic analogue
LaPd$_{2}$Al$_{3}$ becomes superconducting at 0.8 K \cite{06}. The
susceptibility of the Ce-, Pr-, Nd- compounds obeys the
Curie-Weiss law at temperatures above 100 K.

Deviations from the Curie-Weiss law at temperatures below
$\textit{T}$$_{C}$ are apparently due to CF splitting of the
ground-state multiplet \cite{07}, which is also reflected in the
Schottky anomaly observed in specific-heat data collected for
CePd$_{2}$Al$_{3}$ and PrPd$_{2}$Al$_{3}$ \cite{08}.
CePd$_{2}$Al$_{3}$ exhibits heavy-fermion behavior at low
temperatures ($\gamma$ = 380 mJ K$^{-2}$mole$^{-1}$) with a Kondo
effect; \cite{09} nevertheless, it becomes antiferromagnetic below
2.8 K \cite{10}$^{,}$ \cite{11}. PrPd$_{2}$Al$_{3}$ remains
paramagnetic down to 1.5 K. Saturation of susceptibility below 15
K has been attributed to the CF effect yielding the ground-state
singlet \cite{12}. NdPd$_{2}$Al$_{3}$ becomes antiferromagnetic at
$\textit{T}$$_{N}$ = 6.5 K as reported in \cite{08}. The value of
$\textit{T}$$_{N}$ of NdPd$_{2}$Al$_{3}$ has been found to be
varying in the interval from 5.2 to 7.7 K for various samples,
following a linear dependence on the lattice parameter
$\textit{a}$, not on $\textit{c}$, which may be understood with
the exchange interactions inside the $\textit{ab}$ plane
dominating the magnetism in this compound \cite{11}. The most
intriguing case in the $\textit{RE}$Pd$_{2}$Al$_{3}$ family seems
to be SmPd$_{2}$Al$_{3}$. Results published so far on
SmPd$_{2}$Al$_{3}$ have been obtained by measuring only
polycrystals and point to a complex magnetic phase diagram with
several magnetic phase transitions below 12 K \cite{04}. The CF
interaction in SmPd$_{2}$Al$_{3}$ can be reasonably estimated by
considering the well-determined CF parameters for the
NdPd$_{2}$Al$_{3}$ compound \cite{13}$^{,}$ \cite{14} or by using
CF calculations as obtained from first principles, as will be
presented in this paper. Specific-heat measurements of
SmPd$_{2}$Al$_{3}$ have been reported revealing anomalies at 3, 6
and 12 K, indicating magnetic phase transitions \cite{08}. From
susceptibility data three magnetic phase transitions at 4, 4.3 and
12 K \cite{04}, respectively, have been proposed whereas the
electrical resistivity reflects strongly a magnetic phase
transition at 12 K \cite{04}, typical for a transition to
ferromagnetic ordering, the peak-like anomaly around 3.5 K
resembles a transition to a low-temperature antiferromagnetic
phase  \cite{10}. The magnetization curve measured at 2.5 K
indicated two magnetic phase transitions (at approximately 0.3 and
2 T) \cite{10}. It is clear that SmPd$_{2}$Al$_{3}$ presents
unique case of $\textit{4f}$ magnetism and we attempt here to put
this context together with previous results and Sm magnetism
generally.

\section{Experimental and computational details}

The single crystal of SmPd$_{2}$Al$_{3}$ was grown in a triarc
furnace by the Czochralski pulling method from stoichiometric
amounts of elements. The single crystal was a 20-mm long cylinder
with a diameter of 3 - 4 mm. Crystal quality was checked by the
Laue technique. A small part of the crystal was pulverized in an
agate grinding mortar and X-ray powder diffraction (XRPD) data
were recorded on a Seifert powder diffractometer equipped with a
monochromator to provide the Cu K$\alpha$ radiation. The XRPD data
were analyzed by means of the Rietveld profile procedure \cite{15}
using the program FullProf \cite{16}. The final composition of the
crystal was checked by EDX analysis on a FE-SEM Tescan. The Laue
technique was also used for crystal orientation. Samples
appropriate by shape for the magnetization and specific heat
measurements were cut by a fine wire saw (South Bay Technology
Inc, type 810). The sample for the magnetization measurements had
the form of a small beam (1 x 1 x 1.5 mm$^{3}$) with the
rectangular planes oriented perpendicular to the crystallographic
axes $\textit{a}$ and $\textit{c}$, respectively. The
specific-heat samples were small plates (1.5x1.5x0.2 mm$^{3}$)
with the main-plate plane perpendicular to the $\textit{a}$- and
$\textit{c}$-axis, respectively. The orientation of each prepared
sample was subsequently revised again by Laue technique. The
specific-heat, AC susceptibility and magnetization measurements
were performed using the Quantum Design PPMS (Physical Property
Measurement System) and the MPMS (Magnetic Property Measurement
System), respectively. The specific heat was measured at
temperatures from 1.8 to 300 K in magnetic fields up to 9 T. The
magnetization was measured in a temperature range from 1.8 to 400
K and in magnetic fields 0 - 14 T. The AC susceptibility was
measured from 1.8 to 30 K in an AC magnetic field 0.03 mT with a
frequency of 497 Hz. The magnetic field was applied along the
$\textit{a}$- and $\textit{c}$-axis, respectively. To obtain
direct information about the ground-state electronic structure and
magnetic properties, we applied theoretical methods from first
principles. The ground-state electronic structure was calculated
on the basis of density functional theory (DFT) within the local
spin density approximation (LSDA)\cite{17} and generalized
gradient approximation (GGA) \cite{18}. For this purpose, we used
the full potential augmented plane wave plus local-orbitals method
(APW + lo) as implemented in the latest version (WIEN2k) of the
original WIEN code \cite{19}. The calculations were scalar
relativistic and performed with the following parameters:
non-overlapping atomic spheres (AS) radii of 2.8, 2.5 and 2.0 a.u.
(1 a.u. = 52.91 pm) were taken for Sm, Pd and Al, respectively;
the basis for the expansion of the valence electron states (less
than 6 Ry below the Fermi energy) consisted of more than 800 basis
functions (more than 130 APW/atom) plus Sm
($\textit{5s}$,$\textit{5p}$), Pd ($\textit{4p}$) and Al
($\textit{2p}$) local orbitals. The results of our experimental
analysis suggest that the $\textit{4f}$ electrons are localized
atomic-like. Therefore the Sm $\textit{4f}$ states were treated in
an open-core approximation with the stable atomic configuration
$\textit{4f$^{5}$}$. The wave functions in the AS region were
expanded up to \textit{l} = 12 and a density GMAX = 14 was used
for the interstitial charge. Brillouin-zone (BZ) integration was
performed with the tetrahedron method \cite{19} on a 270 special
$\textit{k}$-point mesh (4600$\textit{k}$-points in the full BZ).
We carefully tested the convergence of the results presented with
respect to the parameters mentioned and found them to be fully
sufficient for all the presented characteristics of the
SmPd$_{2}$Al$_{3}$ compound. We also analyzed the temperature
evolution of the magnetic entropy. In addition we included
spin-orbit coupling into the calculations for the electronic
structure of the valence states (delocalized Bloch states); we
found only minor changes in the DOS curves and calculated crystal
field parameters. The calculations first principles of the crystal
field interaction were performed using the method described in
ref. \cite{20}. In these calculations the electronic structure and
corresponding distribution of the ground state charge density was
obtained using the full-potential APW + lo method. The CF
parameters originated from the aspherical part of the total
single-particle DFT potential in the crystal. To eliminate
self-interaction, a self-consistent procedure was first converged
with the $\textit{4f}$ electrons in the core \cite{20}, which was
the open-core approximation used in this work. To diagonalize the
microscopic CF Hamiltonian of the hexagonal point group symmetry,
we used $\textit{J}$ = 5/2, $\textit{J}$ = 7/2 and $\textit{J}$ =
9/2 multiplets and Wybourne parameterization of the CF matrix
elements \cite{21}. Therefore the \textit{J}-mixing and the
intermediate-coupling many-particle $\textit{4f}$-wave functions
were properly taken into account. The CF perturbation matrix had a
dimension of 24 and the resulting eigenenergies and eigenfunctions
were used to calculate the magnetic susceptibility as a function
of temperature.

\section{Results}

We successfully prepared the SmPd$_{2}$Al$_{3}$ single crystal of
high quality as confirmed by the Laue patterns. The XRPD data
contained only the reflections compatible with a hexagonal
structure of the PrNi$_{2}$Al$_{3}$ type(space group
$\textit{P6/mmm}$) with lattice parameters $\textit{a}$ = 5.293
\AA, $\textit{c}$ = 4.064 \AA, which are in good agreement with
published data. The proper 1:2:3 stoichiometry of the grown
crystal was confirmed by EDX analysis within the accuracy of the
method. No spurious phase has been located. We also carried out an
area analysis and concentrations gradients were also studied. No
concentration gradients have been detected.

\textbf{Specific heat}

\begin{figure}
\scalebox{.7}{%
\includegraphics{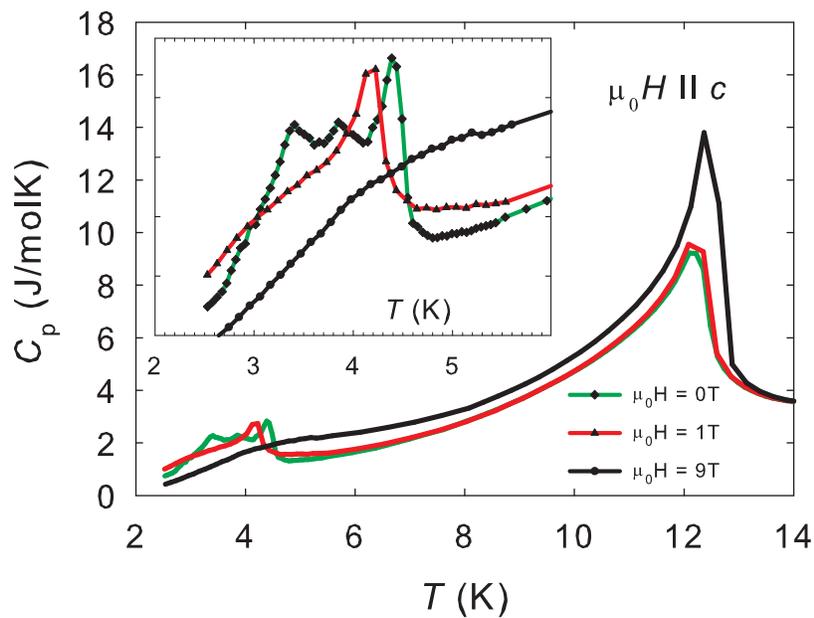}
} \caption{(Color online) Temperature dependence of specific heat
measured in various magnetic fields applied parallel to the
\textit{c}-axis. The inset shows details of the low-temperature
anomalies.}
\label{fig1} 
\end{figure}

\begin{figure}
\scalebox{.7}{%
\includegraphics{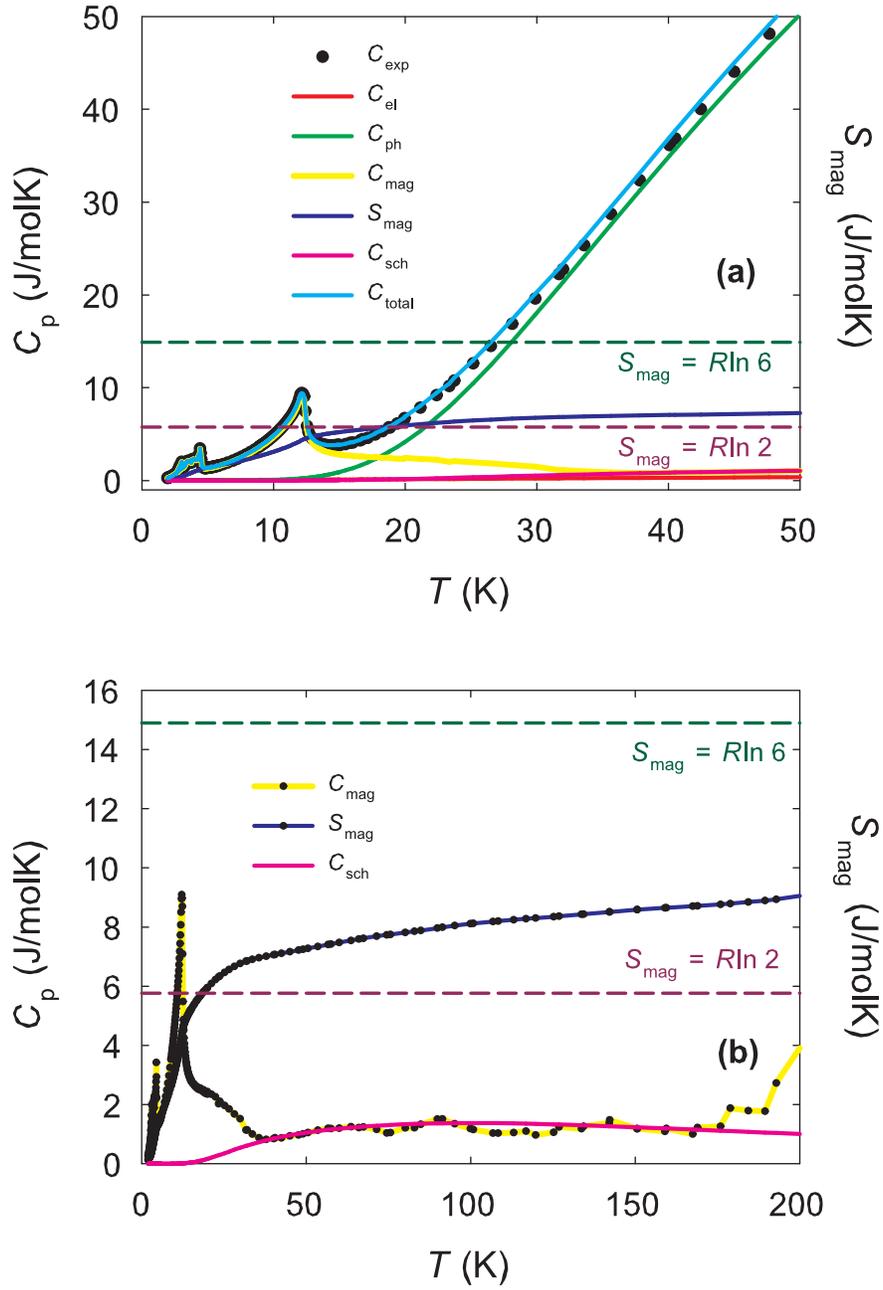}
} \caption{(Color online) Comparative analysis of specific heat:
the graph (\textit{a}) shows the total specific heat and
individual contributions, respectively. Temperature dependence of
magnetic entropy (\textit{b}) is shown, as well.}
\label{fig2}       
\end{figure}

The $\textit{C}$$_{P}$ vs. $\textit{T}$ data were collected by two
series of measurements in a magnetic field applied along the
$\textit{a}$- and $\textit{c}$-axis, respectively. In zero
magnetic field a large and sharp $\textit{l}$-shape anomaly in the
$\textit{C}$$_{P}$($\textit{T}$) dependence, peaking at 12.4 K was
detected and three much smaller peak-like anomalies located at
4.4, 3.9, 3.4 K, respectively (see Fig.\ref{fig1}). We tentatively
denote these temperatures as $\textit{T}$$_{C}$,
$\textit{T}$$_{1}$ , $\textit{T}$$_{2}$ , $\textit{T}$$_{3}$,
respectively. The 12.4-K anomaly became gradually enhanced with
increasing magnetic field applied along the
$\textit{c}$-axis($\textit{B}$$\parallel$$\textit{c}$), though it
remained 'pinned' at nearly the same temperature (it move only
very slightly to higher temperatures). Although such evolution is
rather unusual, the magnetization and resistivity data presented
below confirm that this specific-heat anomaly is apparently
associated with the onset of a ferromagnetic ordering in
SmPd$_{2}$Al$_{3}$ and 12.4 K is the Curie temperature
($\textit{T}$$_{C}$) of this material. The low-temperature
anomalies, which were smeared out in fields
$\textit{B}$$\parallel$$\textit{c}$ higher than 1 T, were
presumably associated with an order-order magnetic phase
transition. The 12.4-K peak remained nearly intact in a magnetic
field up to 9 T applied along the $\textit{a}$-axis (not shown in
the figure), thus confirming a strong magnetocrystalline
anisotropy in the ferromagnetic state. On the other hand, the low
temperature group of anomalies became continuously smeared out
with an increasing field $\textit{B}$$\parallel$$\textit{a}$ (not
shown in the figure). This result indicates that the low
temperature phases ($\textit{T}$ $<$ $\textit{T}$$_{1}$) have
nonzero components (probably antiferromagnetic) in the basal plane
of the hexagonal structure. For proper comparison, in the
following discussion of our results along with existing literature
data on SmPd$_{2}$Al$_{3}$, we have collected together all
existing information (including our results) on the temperatures
of anomalies of specific heat, magnetization in very low fields,
AC susceptibility and electrical resistivity in the
Tab. \ref{tab1}. It is immediately apparent that there is an
unambiguous response of all measured properties to the onset of
magnetic ordering slightly above 12 K. whereas the situation is
not so straightforward in the of features observed between 3 and 5
K. The temperature dependence of the specific heat of
SmPd$_{2}$Al$_{3}$, which is presented over a wide temperature
range in Fig.\ref{fig2}, was considered to be a sum of the
electronic $\textit{C}$$_{el}$, phonon $\textit{C}$$_{ph}$,
magnetic $\textit{C}$$_{mag}$ and Shottky $\textit{C}$$_{Sch}$
contribution, respectively:
\begin{equation}     \label{eq1}
\textit{C$_{p}$} = \textit{C$_{el}$} + \textit{C$_{ph}$} + \textit{C$_{mag}$} + \textit{C$_{Sch}$}
\end{equation}
The most significant is the phonon contribution mainly at higher
temperatures and this contribution was evaluated within the afore
mentioned Debye and Einstein model\cite{22}, the results being
listed in Tab. \ref{tab2}. The values of the phonon contribution
have then been subtracted from the raw specific-heat data. As
input data for the used model, the values of $\gamma$ = 6.0
mJ/molK$^{2}$ \cite{23}, $\textit{T}$$_{D}$ = 180 K \cite{24} and
figures of degeneration of the 4 phonon optical branches 2-5-3-5
\cite{25},  were taken. The remaining values were analyzed by
testing all the free parameters in order to reach the best
agreement with experimental data, based on the equation listed
hereafter for the other specific heat contribution:
\begin{equation}     \label{eq2}
\textit{C$_{ph}$}=R(\frac{1}{1-\alpha_{D}\textit{T}}\textit{$C_{D}$}+\sum_{i=1}^{3N-3}\frac{1}{1-\alpha_{E}\textit{T}}\textit{$C_{E}$})
\end{equation}

\begin{equation}     \label{eq3}
\textit{C$_{el}$}=\textit{$\gamma$T}
\end{equation}

\begin{equation}     \label{eq4}
\textit{C$_{Sch}$}=\frac{R}{\textit{T}^{2}}(\frac{\sum_{i=0}^{n}\Delta_{i}^{2}exp[-\frac{\Delta_{i}}{\textit{T}}]}{\sum_{i=0}^{n}exp[-\frac{\Delta_{i}}{\textit{T}}]}-(\frac{\sum_{i=0}^{n}\Delta_{i}exp[-\frac{\Delta_{i}}{\textit{T}}]}{\sum_{i=0}^{n}exp[-\frac{\Delta_{i}}{\textit{T}}]})^{2})
\end{equation}
The value found for the Sommerfeld coefficient was $\gamma$ = 7
mJ/mol K$^{-2}$, which is a typical figure that may be expected
and measured in Sm intermetallics compounds\cite{23}. We also
analyzed the temperature evolution of the magnetic entropy
(Fig.\ref{fig2}). In the temperature range 1.8 - 20 K we observed
a high jump up to the value \textit{Rln2} and we presume that the
magnetic ground state of our compound is a doublet. An additional
increase of temperature led to a gradual growth of magnetic
entropy. The values for entropy should reach the value
\textit{Rln6} when the energy of the third doublet is achieved,
but unfortunately this doublet is above room temperature. We have
measured specific heat data up to 400 K, but the data were
strongly affected by the transformation of apiezon from
temperatures over 250 K and it was inappropriate to solve the
absolute value of entropy from the experimental data in this
temperature range.

A huge energy gap of 130 K was observed between the ground and the
first excited state, respectively, which is in good agreement with
the findings in \cite{04} and our theoretical calculations. The
existence of a third doublet, indicated to be around room
temperature, motivated us to measure magnetization up to 400 K,
because a strong influence of CF on the susceptibility behavior
was expected, as will be discussed later.

\textbf{AC susceptibility and magnetization}

\begin{figure} 
\scalebox{.7}{%
\includegraphics{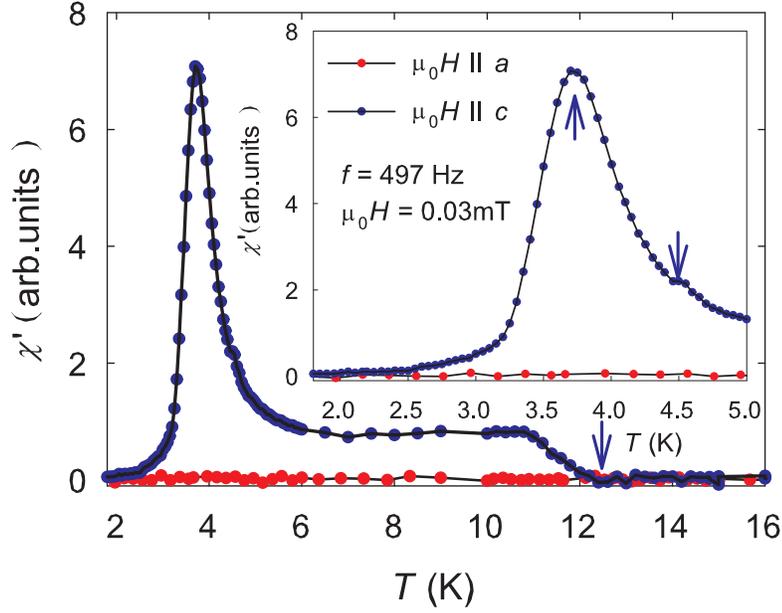}
} \caption{(Color online) Temperature dependence of AC
susceptibility measured with the excitation field applied along
the \textit{a}- and \textit{c}- axis. The inset shows detail at
low temperature and critical temperatures are emphasized by the
arrows.}
\label{fig3}       
\end{figure}

The temperature dependence of the susceptibility measured in the
AC magnetic field applied along the $\textit{a}$- and
$\textit{c}$-axis, respectively, is shown in Fig.\ref{fig3}. The
$\textit{a}$-axis data were characterized by a very low,
nearly-temperature-independent signal free of any considerable
anomaly exceeding signal noise, whereas the data measured in the
AC field applied along the $\textit{c}$-axis exhibited three
anomalies, presumably indicating magnetic phase transitions. When
cooling, the $\textit{c}$-axis data exhibited a sudden upturn
commencing at $\textit{T}$$_{C}$ = 12.4 K, a sharp peak at 3.7 K
and a shoulder at 4.5 K. The latter two features may be
tentatively associated with the magnetic phase transitions at
$\textit{T}$$_{2}$ and $\textit{T}$$_{1}$, respectively, which is
tentatively implied on comparing the data in Tab. \ref{tab1}.

\begin{figure} 
\scalebox{.7}{%
\includegraphics{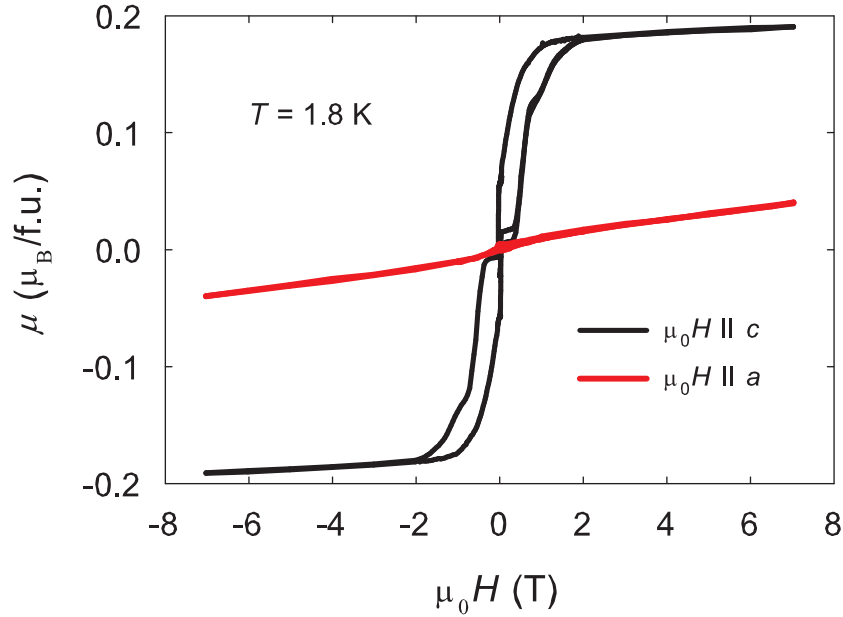}
} \caption{(Color online) Hysteresis loops measured at temperature
1.8 K with applied magnetic field parallel and perpendicular on
the basal plane.}
\label{fig4}       
\end{figure}

\begin{figure} 
\scalebox{.7}{%
\includegraphics{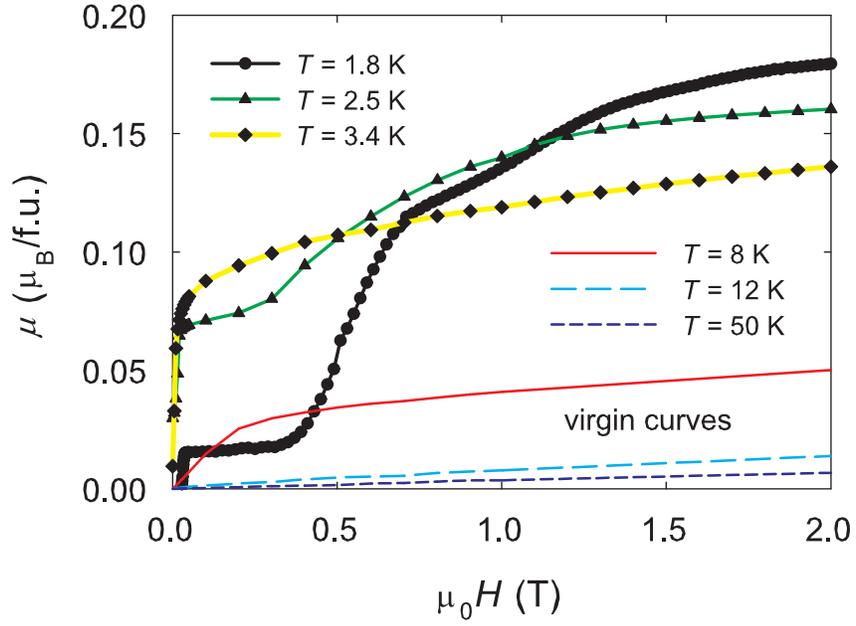}
} \caption{(Color online) The figure shows virgin curves of
SmPd$_{2}$Al$_{3}$ measured along the easy  magnetization axis and
their temperature evolution.}
\label{fig5}       
\end{figure}

The theory \cite{26}, however, says that $\textit{T}$$_{N}$, the
temperature at the maximum of the specific heat and the maximum in
$\partial$($\chi$$\textit{T}$)/$\textit{T}$ vs. $\textit{T}$
frequently does not coincide with the maximum in $\chi$ vs.
$\textit{T}$ in real systems. That was also the case in our study
where the maximum in $\partial$($\chi$$\textit{T}$)/$\textit{T}$
vs. $\textit{T}$ was at 3.4 K, which coincided with the
$\textit{T}$$_{N}$ value determined from specific-heat data. The
dramatic difference between the magnetization curves measured at
1.8 K in the magnetic field applied along the $\textit{a}$- and
$\textit{c}$-axis, respectively, (Fig.\ref{fig4}) corroborated, in
agreement with the specific-heat and AC susceptibility data, the
conclusion that the magnetocrystalline anisotropy in
SmPd$_{2}$Al$_{3}$ was uniaxial and the easy-magnetization
direction was the crystallographic $\textit{c}$-axis. Whereas the
magnetic moment in the field $\textit{B}$$\parallel$$\textit{c}$
is dominant and saturates above 2 T, a very weak paramagnetic-like
linear (in higher fields) response is observed for
$\textit{B}$$\parallel$$\textit{a}$. The slight non-linearity and
hysteresis observed in the loop measured along the hard
magnetization axis was considered to be no more than the result of
an imperfect geometry in the experiment and slight mis-orientation
of the sample than some intrinsic hard axis behavior. The
$\textit{c}$-axis magnetic moment of 0.19 $\mu$$_{B}$/f.u. at 1.8
K observed in a field of 7 T is nearly 5times lower than the
expected ordered moment for the ground-state multiplet of the free
Sm$^{3+}$ ion ($\textit{gJ}$$\mu$$_{B}$ = 0.71 $\mu$$_{B}$), which
is presumably responsible for the entire magnetic behavior of
SmPd$_{2}$Al$_{3}$. We roughly estimated the observed magnetic
moment of the ground state from crystal field calculations. We
obtained a semi-quantitative agreement with the
experimentally-observed magnetization at 1.8 K. For simplicity, we
did not include the exchange interaction due to its complex
nature. The step-like shape of the $\textit{c}$-axis curve in
fields weaker than 2 T was fully reproducible and is apparently
connected with the specific features of samarium magnetism in
SmPd$_{2}$Al$_{3}$. A detailed view of the magnetization loops for
$\textit{B}$$\parallel$$\textit{c}$ is presented in
Fig.\ref{fig5}.

\begin{figure} 
\scalebox{.7}{%
\includegraphics{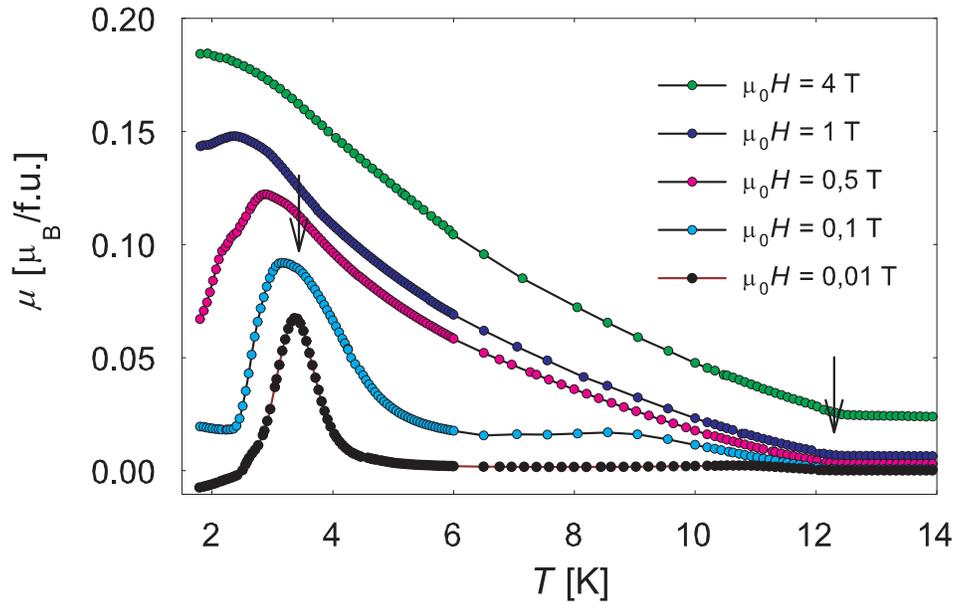}
} \caption{(Color online) Temperature dependence of magnetic
moment of SmPd$_{2}$Al$_{3}$ measured in selected magnetic fields
applied along the \textit{c} with increasing temperature from the
low-temperature ZFC (zero-field cooled) state.}
\label{fig6}       
\end{figure}

\begin{figure} 
\scalebox{.7}{%
\includegraphics{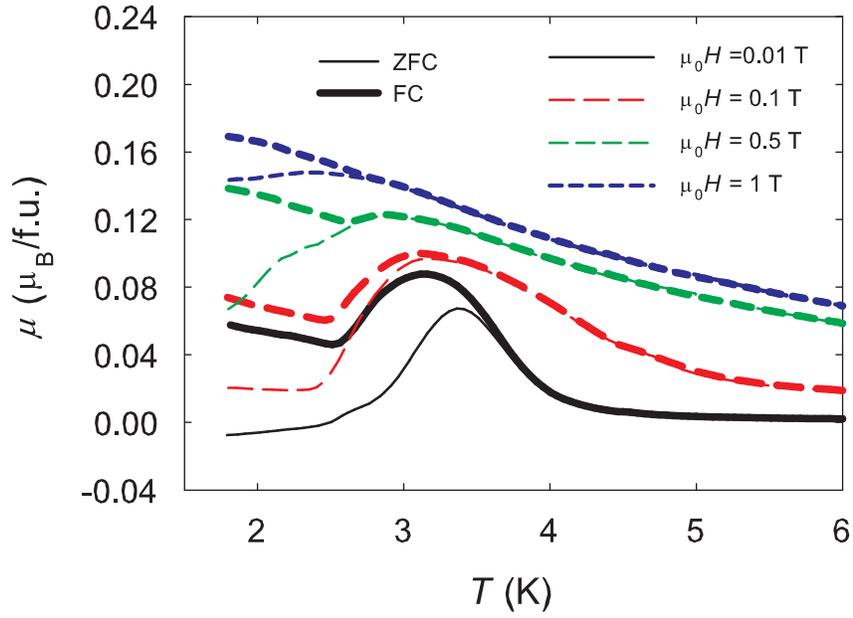}
} \caption{(Color online) FC and ZFC thermomagnetic curves
measured for SmPd$_{2}$Al$_{3}$ measured in selected magnetic
fields applied along \textit{c}.}
\label{fig7}       
\end{figure}

\begin{figure} 
\scalebox{.7}{%
\includegraphics{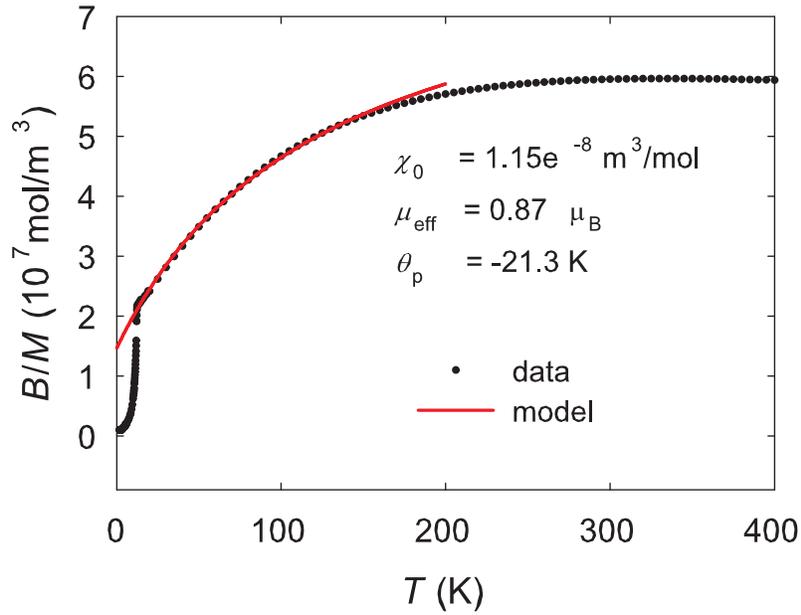}
} \caption{(Color online) Temperature dependence of inverse
susceptibility (\textit{B}/\textit{M}) in magnetic field applied
along the \textit{c}-axis, respectively. The full line represents
the Curie-Weiss fit.}
\label{fig8}       
\end{figure}

A series of four metamagnetic-like transitions can be identified
on the 1.8-K virgin magnetization curve. The onsets of transitions
denoted by arrows can be found at 0.03, 0.35, 0.5 and 0.75 T,
respectively. The transitions in lower fields can still be traced
on the curve measured at 2.5 K, but they have already gone at
temperatures 3.4 K and higher. The magnetization loops measured at
1.8 and 2.5 K exhibited not only some metamagnetic-like features
but also pronounced hysteresis. All these features vanished at
temperatures from 3.4 K. This implies that the ground-state phase,
which forms at $\textit{T}$$_{3}$ is antiferromagnetic and
apparently rather complicated. The magnetization curves measured
at $\textit{T}$$_{3}$ $<$ T $<$ $\textit{T}$$_{C}$, however, still
showed a strong tendency to saturate as usually found in
ferromagnets. The Arrott-plot analysis of the magnetization curves
confirmed the value of $\textit{T}$$_{C}$ = 12.4 K. Above this
temperature, the typical paramagnetic response of the
magnetization to the magnetic field applied parallel to the
$\textit{c}$-axis was observed. In fields parallel to the
$\textit{a}$-axis, the magnetization remained and very weak and
nearly temperature independent below \textit{T}$_{C}$ irrespective
of the dramatic developments in the $\textit{c}$-axis
magnetization. We also measured the temperature dependence of the
magnetic moment in several constant magnetic fields applied along
the $\textit{c}$-axis, the results of which are shown in
Fig.\ref{fig6}. In very low fields the value of the magnetic
moment sharply increased below $\textit{T}$$_{C}$, which was in a
good agreement with the anomaly observed in the specific-heat and
AC-susceptibility data as well as with the evolution of the
magnetization loops with temperature. The 10-mT data were in a
reasonable agreement with the AC susceptibility behavior. The
evolution of 3.4-K anomaly with the increasing magnetic field, as
well as the evolution of the magnetization curves with temperature
decreasing below 5 K, resembled a transition to a low-temperature
antiferromagnetic state, which is stable only in very low magnetic
fields. The bifurcation of the thermomagnetic curve in the ZFC and
FC branches appeared at temperatures below 3.6 K (Fig.\ref{fig7}),
where an hysteresis of the magnetization curves also apeared . The
bifurcation point shifted to lower temperatures with increasing
applied magnetic field. However, an alternative interpretation can
also be considered. It involves a ferromagnetic component to the
magnetic order at the lowest temperature. If all four zero-field
ordering transitions are linked to the field-induced transitions,
then the lowest-field one may be linked to the ferromagnetic
transition at 12.4 K, and would be in fact the coercive field at
which the sample becomes a single domain, as far as the
ferromagnetic component is concerned. The developing difference
(Fig.\ref{fig7}) in zero-field-cooled and field-cooled
magnetization can be considered direct evidence of developing
coercivity, which in turn is linked to ferromagnetism. The upper
three field-induced transitions at 1.8 K may then be directly
connected to the zero-field transitions at 3.4, 3.9 and 4.4 K. The
internal molecular field can be estimated at 1.8 K as 15 T. This
value is roughly comparable with the coercive field. A specific
issue of magnetism in SmPd$_{2}$Al$_{3}$ is the involvement of the
populated CF levels of the first and the second exited multiplet.
This is well documented by the extremely broad Schottky
contribution to the specific heat (see Fig.\ref{fig2}), and, in
addition, a necessary impact on the magnetization (susceptibility)
behavior at elevated temperatures would be expected. The
temperature dependence of the inverse susceptibility
$\textit{B}$/$\textit{M}$ is strongly nonlinear for both field
directions, i.e. it does not follow the Curie-Wiess law. The data
in Fig.\ref{fig8}, however, can be  well fitted by the modified
Curie-Weiss law \cite{27}between 15 and 150 K:

\begin{equation}     \label{eq5}
\chi = \chi_{0} + \frac{C}{\textit{T} - \theta_{P}}
\end{equation}

where the temperature independent term $\chi$$_{0}$ may be
attributed to the temperature independent van Vleck contribution
which is due to the small energy difference between the ground
state multiplet $\textit{J}$ = 5/2 and the first excited multiplet
$\textit{J}$ = 7/2. The fit yields the values for the paramagnetic
Curie temperature $\theta$$_{p}$ = - 21.3 K and effective Sm
moment $\mu$$_{eff}$ = 0.87 $\mu$$_{B}$, which is in a good
agreement with the value of 0.85 $\mu$$_{B}$ expected for the
Sm$^{3+}$ free ion; $\chi$$_{0}$ =  1.15$\cdot$10$^{-8}$
m$^{3}$/mol. The considerable deviation from the fit line at
higher temperatures can be attributed to population of the second
excited multiplet.

\begin{figure} 
\scalebox{.7}{%
\includegraphics{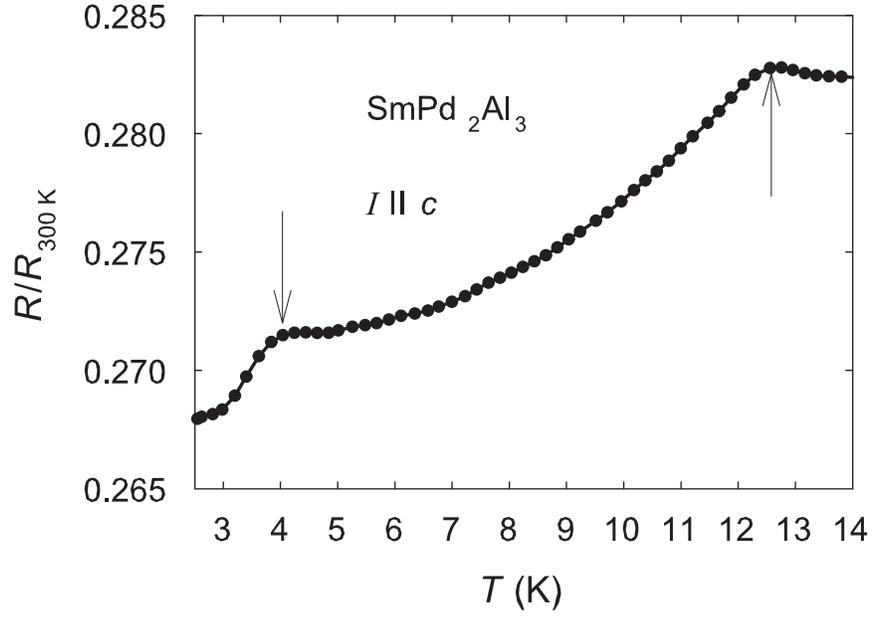}
} \caption{Temperature dependence of relative electrical
resistivity of SmPd$_{2}$Al$_{3}$ with electric current applied
along the \textit{c}-axis in temperature region of magnetic phase
transitions.}
\label{fig9}       
\end{figure}

To see the impact of magnetism on electrical transport we measured
also the temperature dependence of the electrical resistivity for
current along $\textit{c}$ (see Fig.\ref{fig9}). We have detected
a clear cusp around 12.5 K, which is in good agreement to the
onset of ferromagnetism at $\textit{T}$$_{C}$ determined by other
experiments discussed above and also with results presented in the
literature \cite{04}$^{,}$ \cite{10}. Another cusp-like anomaly is
seen around 4 K which coincides with $\textit{T}$$_{2}$.

\textbf{First-principles and CF calculations}

\begin{figure} 
\scalebox{.7}{%
\includegraphics{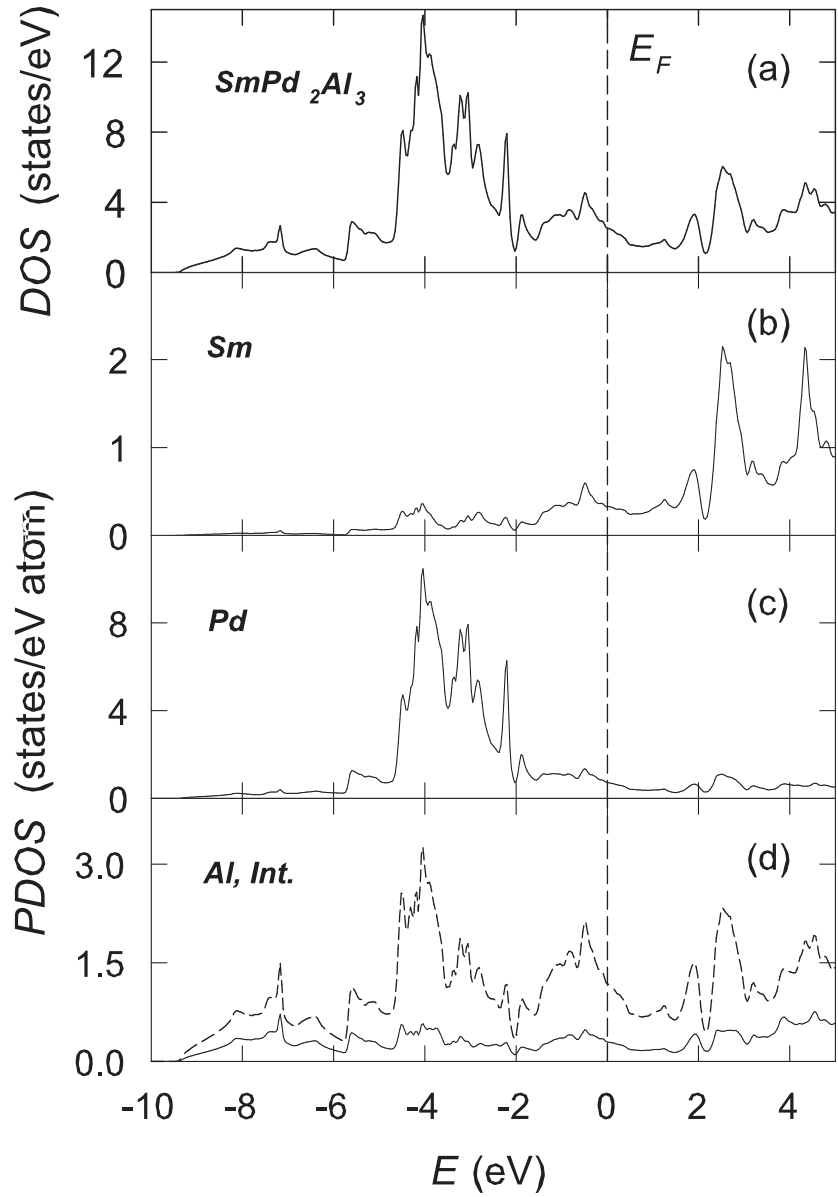}
} \caption{Total DOS (a) and atom-projected DOS (b-d) of
SmPd$_{2}$Al$_{3}$. The projected Sm DOS (b, thick line), Pd (c,
thick line), Al(d, thick line) and the interstitial region (d,
dashed line) are shown. Fermi level is put at zero energy.}
\label{fig10}       
\end{figure}

The calculated GGA electronic density of states (DOS) of
SmPd$_{2}$Al$_{3}$, using the experimentally determined lattice
parameters, is shown in Fig.\ref{fig10}(a). The occupied part of
the DOS has a width of 9.5 eV. The first region, from -9.5 eV to
-4.8 eV, consists mainly of the free electron-like states from the
interstitial region and the Sm-$\textit{6s}$, Pd-$\textit{5s}$,
Al-$\textit{3s}$ and Al-$\textit{3p}$ states  from AS spheres (see
Fig.\ref{fig10} b-d). The following band group from -4.8 eV to the
Fermi level represents mainly the Pd-$\textit{4d}$ states
hybridizing with the Sm-$\textit{5d}$ and Al-$\textit{3p}$ states.
The unoccupied states above the Fermi level have predominantly the
Sm-$\textit{5d}$ character with an admixture of the
Pd-$\textit{4d}$ and Al-$\textit{3p}$ states and the large
contribution from the free electron-like interstitial region
(Fig.\ref{fig10} b-d). The energy position of the localized Sm
$\textit{4f}$$^{5}$ states is correctly below the Fermi level. We
performed also the spin polarized LSDA and GGA calculations in
order to estimate the value of the hybridization induced Pd and Al
magnetic moments ($\textit{4f}$ in core and $\textit{4f}$ in band
as a Bloch states) and found values less than 0.1 $\mu$$_{B}$ and
0.01 $\mu$$_{B}$, respectively. The Fermi level for
SmPd$_{2}$Al$_{3}$ situated at the descending part of the DOS
yielded $\textit{N}$($\textit{E}$$_{F}$) = 2.51 states 1/eV(f.u.).
The orbital analysis of the DOS shows that mainly the
Pd-$\textit{4d}$, Pr-$\textit{5d}$ and Al-$\textit{3p}$ states
that contribute to the total DOS at $\textit{E}$$_{F}$. The value
of the DOS at $\textit{E}$$_{F}$ is too small to cause a
spontaneous magnetic polarization of the Pd-$\textit{4d}$ states.
The value of the DOS at $\textit{E}$$_{F}$ for SmPd$_{2}$Al$_{3}$
corresponds to an electronic specific heat coefficient
$\textit{$\gamma$}$ = 5.9 mJ mol$^{-1}$ K$^{-2}$, which is
somewhat lower than the $\textit{$\gamma$}$ value of 7.0 mJ
mol$^{-1}$ K$^{-2}$ derived from our specific heat data. This
points to a rather low value of the mass-enhancement coefficient
$\lambda$ = 0.18 for SmPd$_{2}$Al$_{3}$ ($\gamma$$_{exp.}$ = (1 +
$\lambda$) $\gamma$$_{band}$), indicating a weak electron-phonon
interaction in the SmPd$_{2}$Al$_{3}$ compound. Next, the magnetic
susceptibility was calculated on the basis of the CF scheme. The
microscopic CF Hamiltonian for the Sm atomic configuration
$\textit{4f}$$^{5}$ in the hexagonal symmetry has four independent
parameters $\textit{$A_{n}^{m}$}$. Reliable CF parameters were
obtained for the NdPd$_{2}$Al$_{3}$ compound by fitting inelastic
neutron data in \cite{13}$^{,}$ \cite{14}; the values obtained
were $\textit{$A_{2}^{0}$}$ = -386 K, $\textit{$A_{4}^{0}$}$ = 42
K, $\textit{$A_{6}^{0}$}$ = 6.8 K and $\textit{$A_{6}^{6}$}$ =
-134 K. At first we decided to check these values by our first
principles calculations.

\begin{figure} 
\scalebox{.65}{%
\includegraphics{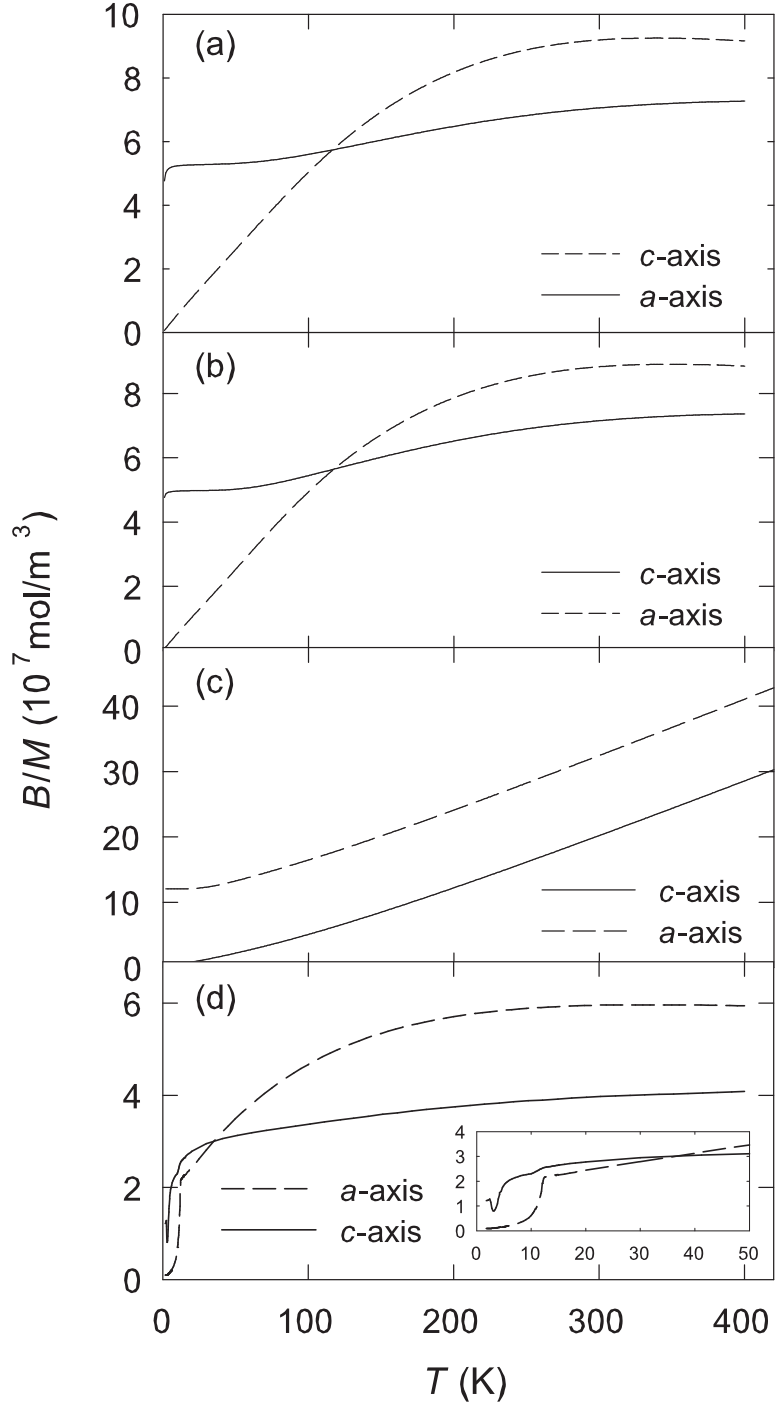}
} \caption{Paramagnetic susceptibility as a function of
temperature of SmPd$_{2}$Al$_{3}$ calculated using CF model (see
text). Calculated curves for the CF parameters from
NdPd$_{2}$Al$_{3}$ (a) and for our first principles calculated CF
parameters for SmPd$_{2}$Al$_{3}$ (b) and calculation the CF
acting on the ground state multiplet \textit{J} = 5/2 only (c).
Experimental data are displayed in the bottom panel (d).}
\label{fig11}       
\end{figure}

The first principles calculation of the crystal field parameters
leads to the following values: $\textit{$A_{2}^{0}$}$ = -307 K,
$\textit{$A_{4}^{0}$}$ = 25 K, $\textit{$A_{6}^{0}$}$ = 1.9 K and
$\textit{$A_{6}^{6}$}$ = -88 K for SmPd$_{2}$Al$_{3}$ using the
GGA form \cite{18} of the exchange-correlation potential. When we
consider spin-orbit interaction the values were:
$\textit{$A_{2}^{0}$}$ = -308, $\textit{$A_{4}^{0}$}$ = 25
$\textit{$A_{6}^{0}$}$ = 1.9 and $\textit{$A_{6}^{6}$}$= -88. The
second-order CF parameter has the correct sign, which determines
the easy $\textit{c}$-axis of SmPd$_{2}$Al$_{3}$ at low
temperatures in agreement with the analysis of our susceptibility
and magnetization data. The other parameters have also the correct
sign and similar values compared to the CF parameters resulting
from the analysis of inelastic neutron data in
\cite{13}$^{,}$\cite{14}. Moreover the calculated gap between the
ground and the first excited doublet is more than 100 K, which is
in goof agreement with the Schottky specific heat analysis and
this fact is also supported by the reliability of both the CF
parameter sets obtained above. Since our theoretical approach is
semi-quantitative only \cite{20}, we may state that a consistent
description of the CF interaction in SmPd$_{2}$Al$_{3}$ compound
is obtained. We would like to emphasize that we calculated a
similar temperature dependence of magnetic susceptibility, which
has an intersection of curves along the $\textit{c}$- and
$\textit{a}$-axis (see Fig.\ref{fig11}), as we found by
experiment. The intersection was not obtained using the CF acting
on the ground state multiplet $\textit{J}$ = 5/2 only so it
requires at least the CF acting on $\textit{J}$ = 5/2,
$\textit{J}$ = 7/2 and $\textit{J}$ = 9/2 quantum subspace. Note
that no intersection has been obtained from the  theory published
by Liu \cite{03}.

A detailed comparison would require a correct description of the
exchange interaction in the paramagnetic region above 12.6 K,
since the simple molecular-field and CF  model was found to
provide only a semiquantitative agreement with our experimental
data, and is therefore definitely too crude for a description of
our measured paramagnetic susceptibility versus temperature data.
Furthermore, we therefore also decided not to fit the measured
data by a too-crude molecular field and CF model approach.

\section{Conclusions}

We have prepared a single crystal of the SmPd$_{2}$Al$_{3}$
compound in order to study specific features of Sm magnetism. The
magnetization, AC susceptibility and specific-heat measurements of
the crystal exposed to magnetic fields applied along the principal
crystallographic axes revealed a strong uniaxial
magnetocrystalline anisotropy (even in paramagnetic state) with a
easy magnetization direction in the $\textit{c}$-axis, which is in
contrast to the easy-plane anisotropy reported for other compounds
in the $\textit{RE}$Pd$_{2}$Al$_{3}$ group. The four magnetic
transitions observed with temperature dependence of the specific
heat at temperatures $\textit{T}$$_{3}$ = 3.4 K,
$\textrm{T}$$_{2}$ = 3.9 K, $\textit{T}$$_{1}$ = 4.4 K and
$\textit{T}$$_{C}$ = 12.4 K, respectively, and detected in part
also in magnetization, AC susceptibility and electrical
resistivity data, point to a complex magnetic phase diagram for
SmPd$_{2}$Al$_{3}$. Although this material becomes ferromagnetic
below $\textit{T}$$_{C}$ = 12.4 K, an antiferromagnetic ground
state seems to become established at low temperatures. The series
of four metamagnetic transitions detected at 0.03, 0.35, 0.5 and
0.75 T, respectively, underlines the complexity  of the Sm
magnetism, which is characterized by a small Sm magnetic moment
and a complex interplay of crystal-field and exchange
interactions. The principal role of CF interaction has been
confirmed by ab initio calculations and our comparison of
calculated and experimental paramagnetic  susceptibility data. To
prove the validity of our scenario on the multiphase magnetic
diagram, as well as on the magnetic phase transitions indicated in
our work suitable microscopic experiments are strongly desired;
namely, neutron diffraction and $\mu$SR spectroscopy. The high
neutron absorption by Sm for a standard thermal neutron wavelength
implies that a hot neutron beam would be necessary for a
successful neutron diffraction experiment with the
SmPd$_{2}$Al$_{3}$ crystal.

\begin{acknowledgements}
This work is part of research plan MSM 0021620834 financed by the
Ministry of Education of the Czech Republic. This work was also
supported by grants of the Czech Science Foundation nos.
202/09/1027 and, 202/09/P354 and the Grant Agency of Charles
University no. 252504.
\end{acknowledgements}

\bibliography{SmPd2Al3J}

\begin{thebibliography}{27}
\expandafter\ifx\csname natexlab\endcsname\relax\def\natexlab#1{#1}\fi
\expandafter\ifx\csname bibnamefont\endcsname\relax
  \def\bibnamefont#1{#1}\fi
\expandafter\ifx\csname bibfnamefont\endcsname\relax
  \def\bibfnamefont#1{#1}\fi
\expandafter\ifx\csname citenamefont\endcsname\relax
  \def\citenamefont#1{#1}\fi
\expandafter\ifx\csname url\endcsname\relax
  \def\url#1{\texttt{#1}}\fi
\expandafter\ifx\csname urlprefix\endcsname\relax\def\urlprefix{URL }\fi
\providecommand{\bibinfo}[2]{#2}
\providecommand{\eprint}[2][]{\url{#2}}

\bibitem[{01()}]{01}
\eprint{G.H. Dieke, Spectra and Energy Levels of Rare Earth Ions in Crystals,
  John Wiley and Sons, 1968, USA}.

\bibitem[{\citenamefont{de~Wijn et~al.}(1967)\citenamefont{de~Wijn, van Diepen,
  and Buschow}}]{02}
\bibinfo{author}{\bibfnamefont{H.}~\bibnamefont{de~Wijn}},
  \bibinfo{author}{\bibfnamefont{A.}~\bibnamefont{van Diepen}},
  \bibnamefont{and} \bibinfo{author}{\bibfnamefont{K.}~\bibnamefont{Buschow}},
  \bibinfo{journal}{Phys.Rev.} \textbf{\bibinfo{volume}{161}},
  \bibinfo{pages}{253} (\bibinfo{year}{1967}).

\bibitem[{\citenamefont{Liu}(2001)}]{03}
\bibinfo{author}{\bibfnamefont{Z.}~\bibnamefont{Liu}},
  \bibinfo{journal}{Phys.Rev. B} \textbf{\bibinfo{volume}{64}},
  \bibinfo{pages}{144407} (\bibinfo{year}{2001}).

\bibitem[{\citenamefont{Kitazawa et~al.}(1993)\citenamefont{Kitazawa, Mori,
  Takano, Yamadaya, Matsushita, and Matsumoto}}]{04}
\bibinfo{author}{\bibfnamefont{H.}~\bibnamefont{Kitazawa}},
  \bibinfo{author}{\bibfnamefont{A.}~\bibnamefont{Mori}},
  \bibinfo{author}{\bibfnamefont{S.}~\bibnamefont{Takano}},
  \bibinfo{author}{\bibfnamefont{T.}~\bibnamefont{Yamadaya}},
  \bibinfo{author}{\bibfnamefont{A.}~\bibnamefont{Matsushita}},
  \bibnamefont{and}
  \bibinfo{author}{\bibfnamefont{T.}~\bibnamefont{Matsumoto}},
  \bibinfo{journal}{Physica B} \textbf{\bibinfo{volume}{186-188}},
  \bibinfo{pages}{661} (\bibinfo{year}{1993}).

\bibitem[{\citenamefont{Geibel et~al.}(1991)\citenamefont{Geibel, Schank,
  Thies, Kitazawa, Bredl, Bohm, Rau, Grauel, Caspary, Helfrich et~al.}}]{05}
\bibinfo{author}{\bibfnamefont{C.}~\bibnamefont{Geibel}},
  \bibinfo{author}{\bibfnamefont{C.}~\bibnamefont{Schank}},
  \bibinfo{author}{\bibfnamefont{S.}~\bibnamefont{Thies}},
  \bibinfo{author}{\bibfnamefont{H.}~\bibnamefont{Kitazawa}},
  \bibinfo{author}{\bibfnamefont{C.}~\bibnamefont{Bredl}},
  \bibinfo{author}{\bibfnamefont{A.}~\bibnamefont{Bohm}},
  \bibinfo{author}{\bibfnamefont{M.}~\bibnamefont{Rau}},
  \bibinfo{author}{\bibfnamefont{A.}~\bibnamefont{Grauel}},
  \bibinfo{author}{\bibfnamefont{R.}~\bibnamefont{Caspary}},
  \bibinfo{author}{\bibfnamefont{R.}~\bibnamefont{Helfrich}},
  \bibnamefont{et~al.}, \bibinfo{journal}{Z Phys. B}
  \textbf{\bibinfo{volume}{84}}, \bibinfo{pages}{1} (\bibinfo{year}{1991}).

\bibitem[{\citenamefont{Zapf et~al.}(2001)\citenamefont{Zapf, Dickey, Freeman,
  Sirvent, and Maple}}]{06}
\bibinfo{author}{\bibfnamefont{V.~S.} \bibnamefont{Zapf}},
  \bibinfo{author}{\bibfnamefont{R.~P.} \bibnamefont{Dickey}},
  \bibinfo{author}{\bibfnamefont{E.~J.} \bibnamefont{Freeman}},
  \bibinfo{author}{\bibfnamefont{C.}~\bibnamefont{Sirvent}}, \bibnamefont{and}
  \bibinfo{author}{\bibfnamefont{M.~B.} \bibnamefont{Maple}},
  \bibinfo{journal}{Phys. Rev. B} \textbf{\bibinfo{volume}{65}},
  \bibinfo{pages}{024437} (\bibinfo{year}{2001}).

\bibitem[{\citenamefont{Mentink et~al.}(1993)\citenamefont{Mentink, Bos,
  Nieuwenhuys, Menovsky, and Mydosh}}]{07}
\bibinfo{author}{\bibfnamefont{S.}~\bibnamefont{Mentink}},
  \bibinfo{author}{\bibfnamefont{N.}~\bibnamefont{Bos}},
  \bibinfo{author}{\bibfnamefont{G.}~\bibnamefont{Nieuwenhuys}},
  \bibinfo{author}{\bibfnamefont{A.}~\bibnamefont{Menovsky}}, \bibnamefont{and}
  \bibinfo{author}{\bibfnamefont{J.}~\bibnamefont{Mydosh}},
  \bibinfo{journal}{Physica B} \textbf{\bibinfo{volume}{186-188}},
  \bibinfo{pages}{497} (\bibinfo{year}{1993}).

\bibitem[{\citenamefont{Ghosh et~al.}(1996)\citenamefont{Ghosh, Ramakrishhnan,
  Chinchure, Marathe, and Chandra}}]{08}
\bibinfo{author}{\bibfnamefont{H.}~\bibnamefont{Ghosh}},
  \bibinfo{author}{\bibfnamefont{S.}~\bibnamefont{Ramakrishhnan}},
  \bibinfo{author}{\bibfnamefont{A.}~\bibnamefont{Chinchure}},
  \bibinfo{author}{\bibfnamefont{V.}~\bibnamefont{Marathe}}, \bibnamefont{and}
  \bibinfo{author}{\bibfnamefont{G.}~\bibnamefont{Chandra}},
  \bibinfo{journal}{Physica B} \textbf{\bibinfo{volume}{223-224}},
  \bibinfo{pages}{354} (\bibinfo{year}{1996}).

\bibitem[{\citenamefont{Suga et~al.}(1993)\citenamefont{Suga, Takeda, Mori,
  Shino, Imada, and Kitazawa}}]{09}
\bibinfo{author}{\bibfnamefont{S.}~\bibnamefont{Suga}},
  \bibinfo{author}{\bibfnamefont{M.}~\bibnamefont{Takeda}},
  \bibinfo{author}{\bibfnamefont{Y.}~\bibnamefont{Mori}},
  \bibinfo{author}{\bibfnamefont{N.}~\bibnamefont{Shino}},
  \bibinfo{author}{\bibfnamefont{S.}~\bibnamefont{Imada}}, \bibnamefont{and}
  \bibinfo{author}{\bibfnamefont{H.}~\bibnamefont{Kitazawa}},
  \bibinfo{journal}{Physica B} \textbf{\bibinfo{volume}{186-188}},
  \bibinfo{pages}{63} (\bibinfo{year}{1993}).

\bibitem[{\citenamefont{Ghosh et~al.}(1993)\citenamefont{Ghosh, Ramakrishnan,
  Malik, and Chandra}}]{10}
\bibinfo{author}{\bibfnamefont{K.}~\bibnamefont{Ghosh}},
  \bibinfo{author}{\bibfnamefont{S.}~\bibnamefont{Ramakrishnan}},
  \bibinfo{author}{\bibfnamefont{S.~K.} \bibnamefont{Malik}}, \bibnamefont{and}
  \bibinfo{author}{\bibfnamefont{G.}~\bibnamefont{Chandra}},
  \bibinfo{journal}{Phys. Rev. B} \textbf{\bibinfo{volume}{48}},
  \bibinfo{pages}{6249} (\bibinfo{year}{1993}).

\bibitem[{\citenamefont{Donni et~al.}(1996)\citenamefont{Donni, Kitazawa,
  Fischer, Vogt, Matsushita, Iimura, and Zolliker}}]{11}
\bibinfo{author}{\bibfnamefont{A.}~\bibnamefont{Donni}},
  \bibinfo{author}{\bibfnamefont{H.}~\bibnamefont{Kitazawa}},
  \bibinfo{author}{\bibfnamefont{P.}~\bibnamefont{Fischer}},
  \bibinfo{author}{\bibfnamefont{T.}~\bibnamefont{Vogt}},
  \bibinfo{author}{\bibfnamefont{A.}~\bibnamefont{Matsushita}},
  \bibinfo{author}{\bibfnamefont{Y.}~\bibnamefont{Iimura}}, \bibnamefont{and}
  \bibinfo{author}{\bibfnamefont{M.}~\bibnamefont{Zolliker}},
  \bibinfo{journal}{Journal of Solid State Chemistry}
  \textbf{\bibinfo{volume}{127}}, \bibinfo{pages}{169} (\bibinfo{year}{1996}).

\bibitem[{\citenamefont{Motoyama et~al.}(2002)\citenamefont{Motoyama, Nishioka,
  and Sato}}]{12}
\bibinfo{author}{\bibfnamefont{G.}~\bibnamefont{Motoyama}},
  \bibinfo{author}{\bibfnamefont{T.}~\bibnamefont{Nishioka}}, \bibnamefont{and}
  \bibinfo{author}{\bibfnamefont{N.}~\bibnamefont{Sato}}, \bibinfo{journal}{J.
  Phys. Soc. Jpn.} \textbf{\bibinfo{volume}{71}}, \bibinfo{pages}{1609}
  (\bibinfo{year}{2002}).

\bibitem[{\citenamefont{Donni et~al.}(1997{\natexlab{a}})\citenamefont{Donni,
  Furrer, Kitazawa, and Zolliker}}]{13}
\bibinfo{author}{\bibfnamefont{A.}~\bibnamefont{Donni}},
  \bibinfo{author}{\bibfnamefont{A.}~\bibnamefont{Furrer}},
  \bibinfo{author}{\bibfnamefont{H.}~\bibnamefont{Kitazawa}}, \bibnamefont{and}
  \bibinfo{author}{\bibfnamefont{M.}~\bibnamefont{Zolliker}},
  \bibinfo{journal}{J. Phys. Condens. Matter} \textbf{\bibinfo{volume}{9}},
  \bibinfo{pages}{5921} (\bibinfo{year}{1997}{\natexlab{a}}).

\bibitem[{\citenamefont{Donni et~al.}(1997{\natexlab{b}})\citenamefont{Donni,
  Furrer, Bauer, Kitazawa, and Zolliker}}]{14}
\bibinfo{author}{\bibfnamefont{A.}~\bibnamefont{Donni}},
  \bibinfo{author}{\bibfnamefont{A.}~\bibnamefont{Furrer}},
  \bibinfo{author}{\bibfnamefont{E.}~\bibnamefont{Bauer}},
  \bibinfo{author}{\bibfnamefont{H.}~\bibnamefont{Kitazawa}}, \bibnamefont{and}
  \bibinfo{author}{\bibfnamefont{M.}~\bibnamefont{Zolliker}},
  \bibinfo{journal}{Z. Phys. B} \textbf{\bibinfo{volume}{104}},
  \bibinfo{pages}{403} (\bibinfo{year}{1997}{\natexlab{b}}).

\bibitem[{\citenamefont{Rietveld}(1969)}]{15}
\bibinfo{author}{\bibfnamefont{H.}~\bibnamefont{Rietveld}},
  \bibinfo{journal}{J. Appl. Cryst.} \textbf{\bibinfo{volume}{2}},
  \bibinfo{pages}{65} (\bibinfo{year}{1969}).

\bibitem[{16()}]{16}
\eprint{http://www.ill.eu/sites/fullprof/}.

\bibitem[{\citenamefont{Perdew and Wang}(1992)}]{17}
\bibinfo{author}{\bibfnamefont{J.~P.} \bibnamefont{Perdew}} \bibnamefont{and}
  \bibinfo{author}{\bibfnamefont{Y.}~\bibnamefont{Wang}},
  \bibinfo{journal}{Phys. Rev. B} \textbf{\bibinfo{volume}{45}},
  \bibinfo{pages}{13244} (\bibinfo{year}{1992}).

\bibitem[{\citenamefont{Perdew et~al.}(1996)\citenamefont{Perdew, Burke, and
  Ernzerhof}}]{18}
\bibinfo{author}{\bibfnamefont{J.~P.} \bibnamefont{Perdew}},
  \bibinfo{author}{\bibfnamefont{K.}~\bibnamefont{Burke}}, \bibnamefont{and}
  \bibinfo{author}{\bibfnamefont{M.}~\bibnamefont{Ernzerhof}},
  \bibinfo{journal}{Phys. Rev. Lett.} \textbf{\bibinfo{volume}{77}},
  \bibinfo{pages}{3865} (\bibinfo{year}{1996}).

\bibitem[{19()}]{19}
\eprint{P. Blaha and K. Schwarz and G. Madsen and D. Kvasnicka and J. Luitz,
  www.wien2k.at, 2001, ISBN 3-9501031-1-2}.

\bibitem[{\citenamefont{Divi\v{s} et~al.}(2005)\citenamefont{Divi\v{s}, Rusz,
  Michor, Hilscher, Blaha, and Schwarz}}]{20}
\bibinfo{author}{\bibfnamefont{M.}~\bibnamefont{Divi\v{s}}},
  \bibinfo{author}{\bibfnamefont{J.}~\bibnamefont{Rusz}},
  \bibinfo{author}{\bibfnamefont{H.}~\bibnamefont{Michor}},
  \bibinfo{author}{\bibfnamefont{G.}~\bibnamefont{Hilscher}},
  \bibinfo{author}{\bibfnamefont{P.}~\bibnamefont{Blaha}}, \bibnamefont{and}
  \bibinfo{author}{\bibfnamefont{K.}~\bibnamefont{Schwarz}},
  \bibinfo{journal}{J. Alloys Comp.} \textbf{\bibinfo{volume}{29}},
  \bibinfo{pages}{403} (\bibinfo{year}{2005}).

\bibitem[{\citenamefont{Nekvasil et~al.}(1995)\citenamefont{Nekvasil,
  Divi\v{s}, Hilscher, and Holland-Moritz}}]{21}
\bibinfo{author}{\bibfnamefont{V.}~\bibnamefont{Nekvasil}},
  \bibinfo{author}{\bibfnamefont{M.}~\bibnamefont{Divi\v{s}}},
  \bibinfo{author}{\bibfnamefont{G.}~\bibnamefont{Hilscher}}, \bibnamefont{and}
  \bibinfo{author}{\bibfnamefont{E.}~\bibnamefont{Holland-Moritz}},
  \bibinfo{journal}{J. Alloys Comp.} \textbf{\bibinfo{volume}{225}},
  \bibinfo{pages}{578} (\bibinfo{year}{1995}).

\bibitem[{\citenamefont{Svoboda et~al.}(2001)\citenamefont{Svoboda,
  Javorsk\'{y}, Divi\v{s}, Sechovsk\'{y}, Honda, Oomi, and Menovsky}}]{22}
\bibinfo{author}{\bibfnamefont{P.}~\bibnamefont{Svoboda}},
  \bibinfo{author}{\bibfnamefont{P.}~\bibnamefont{Javorsk\'{y}}},
  \bibinfo{author}{\bibfnamefont{M.}~\bibnamefont{Divi\v{s}}},
  \bibinfo{author}{\bibfnamefont{V.}~\bibnamefont{Sechovsk\'{y}}},
  \bibinfo{author}{\bibfnamefont{F.}~\bibnamefont{Honda}},
  \bibinfo{author}{\bibfnamefont{G.}~\bibnamefont{Oomi}}, \bibnamefont{and}
  \bibinfo{author}{\bibfnamefont{A.~A.} \bibnamefont{Menovsky}},
  \bibinfo{journal}{Phys. Rev. B} \textbf{\bibinfo{volume}{63}},
  \bibinfo{pages}{212408} (\bibinfo{year}{2001}).

\bibitem[{\citenamefont{Vejpravov\'{a}
  et~al.}(2003)\citenamefont{Vejpravov\'{a}, Svoboda, Rotter, Doerr, and
  Loewenhaupt}}]{23}
\bibinfo{author}{\bibfnamefont{J.}~\bibnamefont{Vejpravov\'{a}}},
  \bibinfo{author}{\bibfnamefont{P.}~\bibnamefont{Svoboda}},
  \bibinfo{author}{\bibfnamefont{M.}~\bibnamefont{Rotter}},
  \bibinfo{author}{\bibfnamefont{M.}~\bibnamefont{Doerr}}, \bibnamefont{and}
  \bibinfo{author}{\bibfnamefont{M.}~\bibnamefont{Loewenhaupt}},
  \bibinfo{journal}{Physica B} \textbf{\bibinfo{volume}{329-333}},
  \bibinfo{pages}{504} (\bibinfo{year}{2003}).

\bibitem[{\citenamefont{Schwerin et~al.}(1996)\citenamefont{Schwerin, Becker,
  Eichler, and Mydosh}}]{24}
\bibinfo{author}{\bibfnamefont{M.}~\bibnamefont{Schwerin}},
  \bibinfo{author}{\bibfnamefont{B.}~\bibnamefont{Becker}},
  \bibinfo{author}{\bibfnamefont{A.}~\bibnamefont{Eichler}}, \bibnamefont{and}
  \bibinfo{author}{\bibfnamefont{J.}~\bibnamefont{Mydosh}},
  \bibinfo{journal}{J. Magn. Magn. Materials}
  \textbf{\bibinfo{volume}{226-230}}, \bibinfo{pages}{176}
  (\bibinfo{year}{1996}).

\bibitem[{\citenamefont{Svoboda et~al.}(2006)\citenamefont{Svoboda,
  Vejpravov\'{a}, Dani\v{s}, and Mihalik}}]{25}
\bibinfo{author}{\bibfnamefont{P.}~\bibnamefont{Svoboda}},
  \bibinfo{author}{\bibfnamefont{J.}~\bibnamefont{Vejpravov\'{a}}},
  \bibinfo{author}{\bibfnamefont{S.}~\bibnamefont{Dani\v{s}}},
  \bibnamefont{and} \bibinfo{author}{\bibfnamefont{M.}~\bibnamefont{Mihalik}},
  \bibinfo{journal}{Physica B} \textbf{\bibinfo{volume}{378-380}},
  \bibinfo{pages}{1107} (\bibinfo{year}{2006}).

\bibitem[{\citenamefont{Fisher}(1962)}]{26}
\bibinfo{author}{\bibfnamefont{M.}~\bibnamefont{Fisher}},
  \bibinfo{journal}{Phil. Mag.} \textbf{\bibinfo{volume}{7}},
  \bibinfo{pages}{1731} (\bibinfo{year}{1962}).

\bibitem[{\citenamefont{El-Hagary et~al.}(2005)\citenamefont{El-Hagary, Michor,
  and Hilscher}}]{27}
\bibinfo{author}{\bibfnamefont{M.}~\bibnamefont{El-Hagary}},
  \bibinfo{author}{\bibfnamefont{M.}~\bibnamefont{Michor}}, \bibnamefont{and}
  \bibinfo{author}{\bibfnamefont{G.}~\bibnamefont{Hilscher}},
  \bibinfo{journal}{Physica B} \textbf{\bibinfo{volume}{284-288}},
  \bibinfo{pages}{1489} (\bibinfo{year}{2005}).

\end{thebibliography}

\pagebreak

\begin{table}
\caption{ Summary of temperatures of anomalies in specific heat,
AC susceptibility, low-field magnetization and electrical
resistivity observed in our results in comparison with the
parameters of magnetic transitions in SmPd$_{2}$Al$_{3}$ reported
in the literature.}
\begin{tabular}{cccccc}
\hline \hline
property & Source & \multicolumn{4}{c}Anomaly location [K]\\
\hline
 & & \textit{T$_{C}$} &  \textit{T$_{1}$} & \textit{T$_{2}$} & \textit{T$_{3}$}\\
\hline
\textit{$C_{p}$} & [this work] & 12.4 & 4.4 & 3.9 & 3.4  \\
\textit{$C_{p}$} & \cite{04} & 12.5 & 4.3 & 4 & - \\
\textit{C$_{p}$} & \cite{08} & 12.5 & 4.7 & - & 3.7 \\
$\chi$$_{AC}$ & [this work] & 12.5 & 4.5 & - & 3.7/3.4  \\
\textbf{$\textit{M}$} & [this work] & 12.3 & 4.4 & - & 3.4 \\
\textit{M} & \cite{10} & 12.5 & - & 4 & 3.3 \\
\textit{M} & \cite{04} & 12.5 & 4.3 & 4 & -  \\
$\rho$ & [this work] & 12.4 & 4.4 & - & 3.4  \\
$\rho$ & \cite{10}  & 12.5 & 4.5 & - & 3.6 \\
$\rho$ & \cite{04}  & 12.0 & - & - & - \\
\hline\hline
\end{tabular}
\label{tab1}
\end{table}

\begin{table}
\caption{ The values and degeneracy of branches of the phonon
contribution to the total specific heat evaluated on the basis of
Einstein and Debye models.}
\begin{tabular}{cccc}
\hline \hline
Branches & Degeneracy & $\theta$ [K]& $\alpha$ [10$^{-4}$K$^{-1}$]\\
\hline
$\theta$$_{D}$ &  & 191 & 5.0  \\
$\theta$$_{E1}$ & 2 & 103 & 5.0 \\
$\theta$$_{E2}$ & 5 & 210 & 4.5 \\
$\theta$$_{E3}$ & 3 & 154 & 4.0 \\
$\theta$$_{E4}$ & 5 & 420 & 3.0 \\
\hline\hline
\end{tabular}
\label{tab2}
\end{table}

\begin{table}
\caption{ Energy of the 3 Kramer's doublets after CF splitting of
the degenerated ground state $\textit{J}$ = 5/2 multiplet.}
\begin{tabular}{cc}
\hline \hline
Levels & Energy [K]\\
\hline
$\Delta$$_{1}$ & 0 \\
$\Delta$$_{2}$ & 130 $\pm$ 15 \\
$\Delta$$_{3}$ & 300 $\pm$ 40 \\
\hline\hline
\end{tabular}
\label{tab3}
\end{table}

\end{document}